\documentclass[runningheads]{llncs} 

\usepackage[utf8]{inputenc}
\usepackage{graphicx}
\usepackage{float}
\usepackage{import}
\usepackage[linesnumbered,ruled]{algorithm2e}
\SetKwComment{Comment}{/* }{ */}
\usepackage{tabularx}
\usepackage{xcolor}
\usepackage{amsmath}
\usepackage{hyperref}
\graphicspath{{./Images/}}

\usepackage{tikz}
\usepackage{hyperref}

\newcommand{\orcid}[1]{%
    \href{https://orcid.org/#1}{\includegraphics[scale=0.08]{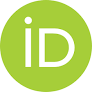}}%
}
\begin{document}

\title{Direction Estimation of Sound Sources Using Microphone Arrays and Signal Strength}

\author{
Mahdi Ali Pour\orcid{0000-0002-3860-020X}\inst{1} \and
Zahra Habibzadeh\orcid{0009-0008-2608-4851}\inst{2}
}

\institute{
Computer Science \& Engineering, Sabancı University, Faculty of Engineering and Natural Science, Istanbul, Turkey \\
\email{mahdialipour@sabanciuniv.edu}
\and
Artificial Intelligence and Robotics, University of Tehran, School of Electrical and Computer Engineering, Tehran, Iran \\
\email{z.habibzadeh213@ut.ac.ir}
}

\titlerunning{Direction Estimation of Sound Sources}
\authorrunning{Mahdi Ali Pour et al.}

\maketitle


\begin{abstract}
Sound-tracking refers to the process of determining the direction from which a sound originates, making it a fundamental component of sound source localization. This capability is essential in a variety of applications, including security systems, acoustic monitoring, and speaker tracking, where accurately identifying the direction of a sound source enables real-time responses, efficient resource allocation, and improved situational awareness. While sound-tracking is closely related to localization, it specifically focuses on identifying the direction of the sound source rather than estimating its exact position in space. Despite its utility, sound-tracking systems face several challenges, such as maintaining directional accuracy and precision, along with the need for sophisticated hardware configurations and complex signal processing algorithms. This paper presents a sound-tracking method using three electret microphones. We estimate the direction of a sound source using a lightweight method that analyzes signals from three strategically placed microphones. By comparing the average power of the received signals, the system infers the most probable direction of the sound. The results indicate that the power level from each microphone effectively determines the sound source direction. The proposed technique offers distinct advantages over existing sound-tracking methods. Our system employs a straightforward and cost-effective hardware design, ensuring simplicity and affordability in implementation. It achieves a localization error of less than 6 degrees and a precision of 98\%. Additionally, its effortless integration with various systems makes it versatile and adaptable. Consequently, this technique presents a robust and reliable solution for sound-tracking and localization, with potential applications spanning diverse domains such as security systems, smart homes, and acoustic monitoring.

The source code of the project is available on our GitHub repository: \url{https://github.com/mahdi943/soundlocalization}
\end{abstract}

\section{Introduction}

The ability to detect sounds and determine their direction has become important in modern intelligent systems. This functionality supports various applications, such as smart home automation and interactive voice-based systems. For instance, identifying the direction of a sound can help locate a speaker in a room, trace the source of environmental noise, or respond appropriately to specific acoustic events \cite{ijerph18179039}. As demand grows for systems that are more responsive to their surroundings, directional sound sensing has the potential as a key component in enhancing situational awareness and context-aware behavior.

Technology designed to locate and follow sound sources has been widely integrated into robotics \cite{desai2022review}, search-and-rescue operations \cite{deleforge2020drone}, and surveillance \cite{KHAN2025100313}, demonstrating its effectiveness in real-world scenarios \cite{9079214}. Such advancements enable more intuitive interactions with devices, improving responsiveness to user commands. We believe directional audio detection is particularly beneficial in applications where pinpointing a sound's origin enhances functionality. For instance, in smart homes, voice assistants like Amazon Alexa and Google Assistant utilize this capability to determine a user's location within a room, allowing for more effective responses. Similarly, adaptive audio systems in home theaters use spatial awareness to enhance sound distribution, creating immersive listening experiences tailored to the user's position.

While sound localization estimates the exact position of a sound source, sound tracking focuses on determining its direction. Localization typically requires complex sensor arrays and heavy computation, whereas tracking provides directional information in real time with fewer sensors and simpler processing, offering a lightweight, cost-effective alternative when full coordinates are unnecessary.

Many existing methods rely on complex hardware and intensive signal processing, which limits their practicality for low-cost applications. Our approach demonstrates that reliable direction estimation can be achieved using inexpensive, widely available hardware, without high-end processors or elaborate sensor arrays. This makes it suitable for educational, prototyping, and resource-constrained environments.

This work focuses on enabling directional sound-tracking for responsive systems such as smart homes, robots, and compact automation tools. By estimating sound direction in real time, the method supports adaptive behavior in resource-constrained environments as a lightweight alternative to computationally intensive localization techniques.

Our approach uses a vector-based method to compute the sum of microphone signal vectors, allowing accurate direction detection with minimal hardware. This capability is valuable for security, automation, and smart devices that benefit from rapid and reliable sound tracking.

Unlike machine learning or beamforming approaches, our method emphasizes simplicity and affordability. Using readily available Arduino boards and electret microphones, we lower hardware costs and configuration effort, making sound-tracking accessible for budget-sensitive or rapid-deployment applications.

\section{Literture Review}
\subsection{Sound Source Localization (SSL)}

Sound source localization (SSL) is a core component in many applications, including source separation \cite{chazan2019multi}, automatic speech recognition (ASR) \cite{7497454}, speech enhancement \cite{xenaki2018sound}, noise control \cite{chiariotti2019acoustic}, and room acoustics analysis \cite{amengual2017spatial}. SSL typically aims to determine the position or direction of a sound source using sensor data.

Classical approaches to SSL include Time Difference of Arrival (TDOA), which computes the delay in signal arrival between multiple microphones, and phase-based techniques that leverage phase shifts of the audio signal across sensors \cite{wang2024sound,heydari2023real}. Beamforming methods also play a central role in SSL, using microphone arrays to estimate direction based on signal alignment over time \cite{1323088,lee2005real}.

Other localization strategies involve Bayesian filtering \cite{valin2006robust,nguyen2016localizing}, triangulation \cite{gabriel20192d,ishi2015speech}, and direction-surface intersection models \cite{9025940,hoshiba2017design}, all offering increased precision at the cost of complexity.

Microphone arrays are frequently employed for their spatial resolution capabilities, although systems with large arrays or high sampling rates can become computationally intensive \cite{portello2014active}. Recent research has explored machine learning approaches to classify sound direction using raw or pre-processed signal data \cite{rabbi2011passive,park2014sensing}.

In contrast to these hardware-heavy or computationally demanding approaches, our work emphasizes a lightweight solution that leverages low-cost, readily available components. This novelty allows us to achieve directional estimation without the complexity and resource requirements of traditional methods.

\subsection{Sound Source Tracking}

In dynamic environments, tracking the direction of a moving sound source over time is essential. Traditional SSL methods have been extended to enable real-time tracking using Kalman filters, particle filters, or hybrid probabilistic models \cite{lauzon2017localization}. These models can update the sound source’s estimated direction continuously, improving robustness to noise and source movement.

Sound-based sensing also plays a significant role in human-centered applications, such as speaker tracking for smart environments \cite{li2016reverberant}, productivity tools \cite{tan2013sound,lee2013sociophone}, and acoustic monitoring in mobile and wearable devices \cite{rabbi2011passive}.

One relevant example is intensity-based tracking, where the direction is inferred by comparing sound intensity differences across microphones \cite{10877355}. While less complex than TDOA or beamforming, these methods may offer faster processing at the expense of some spatial accuracy.

\subsection{Summary and Motivation}

While traditional SSL and tracking methods offer high accuracy, they often require large microphone arrays, complex signal processing algorithms, and high computational power—making them less ideal for lightweight or low-cost systems.

Instead, our work focuses on a simple, intensity-based method that relies on only three microphones and does not require time delay or phase computations. This enables a more practical, scalable, and cost-effective solution for indoor directional sound-tracking, suitable for applications such as smart homes, speaker-following systems, and acoustic event detection.

\section{Methodology}

In this section, we explain how our approach determines the direction of a sound source by analyzing the intensity of sound signals captured by three microphones. The microphones are placed at 0°, 120°, and 240°, measured counterclockwise from a defined 0° reference axis aligned with the x-axis of the device. When the sound intensity detected by any of the microphones exceeds a predefined threshold (Algorithm \ref{alg:finder2}, line 3), the system begins collecting a fixed number of amplitude samples from all three microphones (line 6). These samples reflect the sound signal's strength over time and are used in the subsequent direction estimation process.

\begin{algorithm}[t]
\caption{Sound Localization Algorithm}
\label{alg:finder2}
\textbf{Input}: Sound signal \( S \), number of samples \( N \), threshold \( T \)\ \\
\textbf{Output}: Sound angle, sound magnitude, servo motor position

\If{\( S > T \)}
{
    \ForEach{microphone \( M_i \)}{
        \textit{// Collect and average samples}\\
        \( Array_i \gets \) Collect \( N \) samples from \( S \)\\
        \( Avg_i \gets \) Average of \( Array_i \)\\

        \textit{// Convert average signals to polar form}\\
        \( V_i \gets \) Polar vector from \( Avg_i \)\\
        \( R_i \gets \) Rectangular form of \( V_i \)\\
    }

    \textit{// Calculate the resultant vector}\\
    \( X \gets \sum R_i[0] \) \quad \textit{// Sum of x-components}\\
    \( Y \gets \sum R_i[1] \) \quad \textit{// Sum of y-components}\\

    \textit{// Compute final magnitude and angle}\\
    \( ResultMag \gets \sqrt{X^2 + Y^2} \)\\
    \( ResultAngle \gets \text{atan2}(Y, X) \)\\

    \textit{// Showing the position with servo motor }\\
    \( Servo \gets ServoPos \)
}
\end{algorithm}

For each microphone, the average amplitude of the collected samples is calculated (line 7). These averages are then interpreted as vectors in polar form, where the magnitude corresponds to the average signal power received, and the angle corresponds to the microphone’s position. Specifically, we define three polar vectors, each represented as a tuple in the form $(\theta, r)$, where $\theta$ is the angle and $r$ is the average signal power. These vectors—$\alpha$, $\beta$, and $\gamma$—are shown in Equations \ref{ave}–\ref{ave3}.

\begin{equation} \label{ave} \alpha = (120^\circ, \text{PowerAvg}_1) \end{equation} \begin{equation} \label{ave2} \beta = (0^\circ, \text{PowerAvg}_2) \end{equation} \begin{equation} \label{ave3} \gamma = (240^\circ, \text{PowerAvg}_3) \end{equation}

These polar vectors are converted by Equations \ref{rect}, \ref{rect2} and \ref{rect3} to rectangular (Cartesian) form using standard trigonometric conversion, resulting in vectors \( R = (x, y) \), where \( x = r \cdot \cos(\theta) \) and \( y = r \cdot \sin(\theta) \).

\begin{equation}
\label{rect}
R_1 = (r_\alpha \cdot \cos(\theta_\alpha),\ r_\alpha \cdot \sin(\theta_\alpha))
\end{equation}

\begin{equation}
\label{rect2}
R_2 = (r_\beta \cdot \cos(\theta_\beta),\ r_\beta \cdot \sin(\theta_\beta))
\end{equation}

\begin{equation}
\label{rect3}
R_3 = (r_\gamma \cdot \cos(\theta_\gamma),\ r_\gamma \cdot \sin(\theta_\gamma))
\end{equation}

The three rectangular vectors are summed component-wise \( R_{\text{total}} = (X, Y) \) to compute the resulting direction vector, as defined in Equations \ref{sum} and \ref{sum2}.

\begin{equation}
\label{sum}
X = \sum_{i=1}^{3} R_{x_i}
\end{equation}

\begin{equation}
\label{sum2}
Y = \sum_{i=1}^{3} R_{y_i}
\end{equation}

The final direction of the sound source is determined by converting the resultant rectangular vector back to polar coordinates. 
The magnitude and angle are computed as shown in Equations~\ref{res} and~\ref{res2}.

\begin{equation}
\label{res}
\text{ResultMag} = \sqrt{X^2 + Y^2}
\end{equation}

\begin{equation}
\label{res2}
\text{ResultAngle} = \text{atan2}(Y, X)
\end{equation}

The \texttt{atan2} function returns the angle (in radians) between the positive \( x \)-axis and the point \( (X, Y) \), accounting for the correct quadrant of the angle. This ensures an accurate directional result in the full \([-\pi, \pi]\) range.

\begin{figure}[H]
  \centering
  \begin{minipage}{0.4\textwidth}
    \centering
    \includegraphics[scale=0.4]{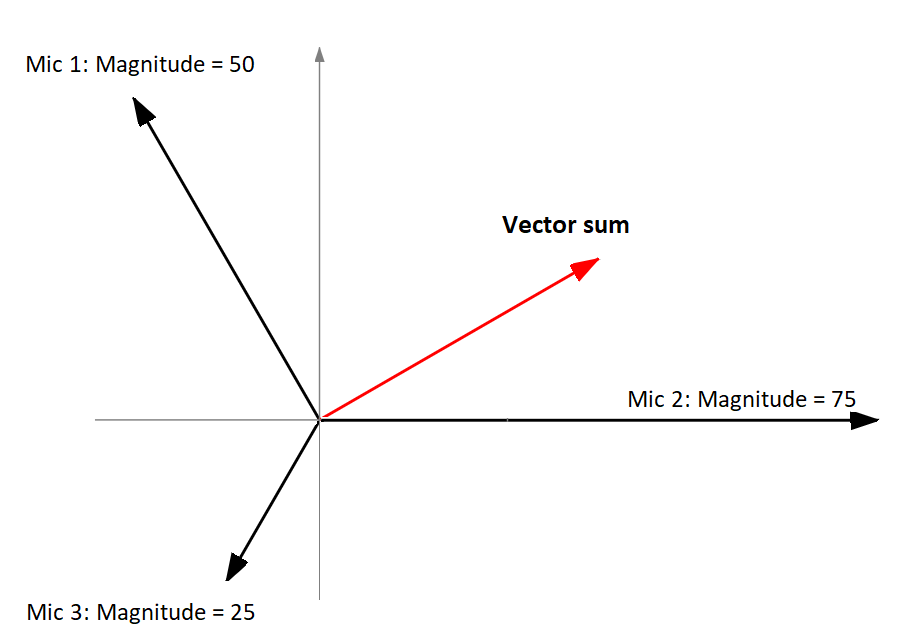}
    \caption{Example showing the magnitudes of signals from three different microphones in black, and their vector sum in red, representing the resulting sound direction.}
    \label{fig:trace}
  \end{minipage}\hfill
  \begin{minipage}{0.5\textwidth}
    \centering
    \includegraphics[scale=0.09]{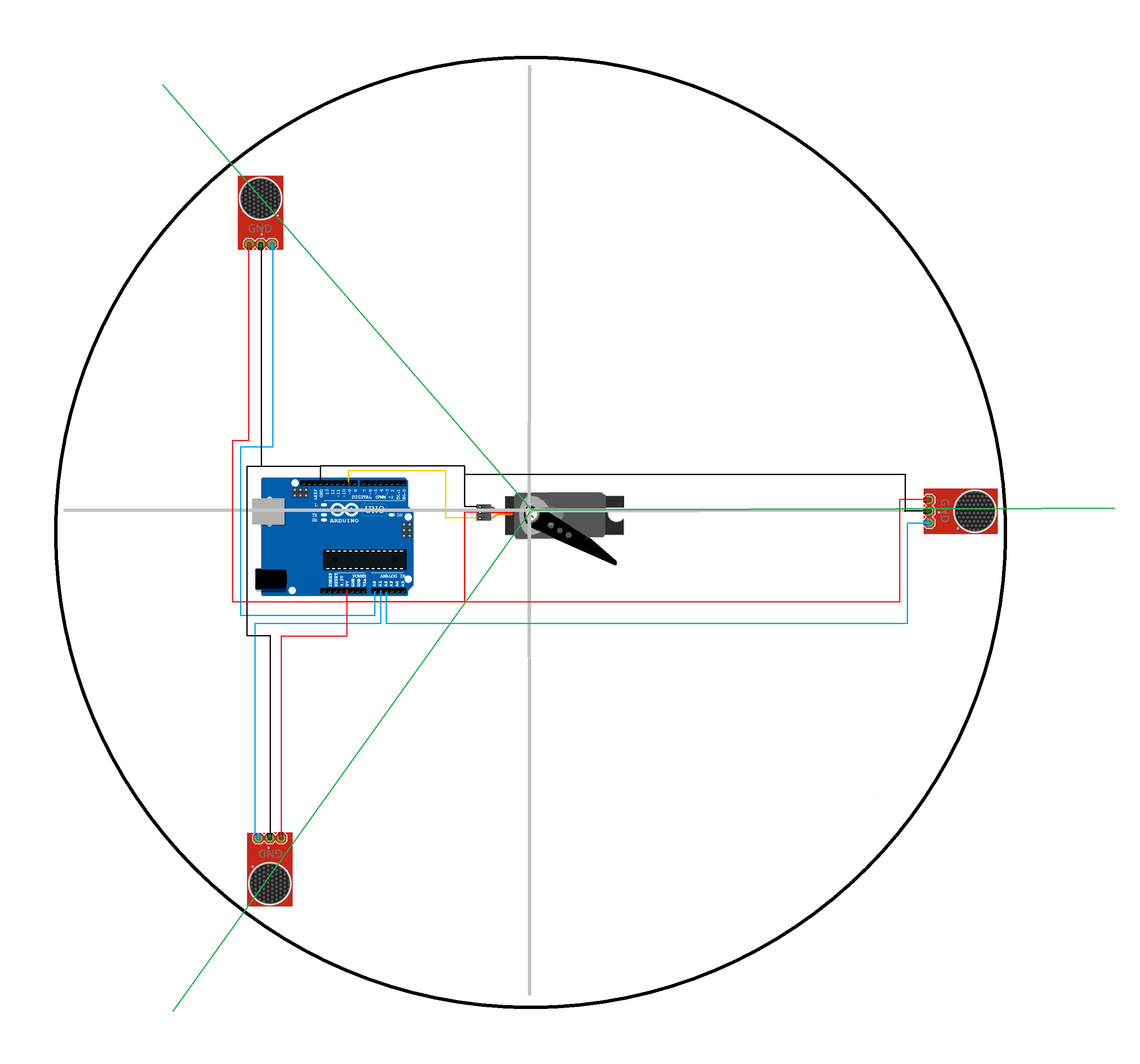}
    \caption{Circuit diagram represents the location of microphones and the servo motor in the center. Each microphone is 120 degrees from the next microphone, which shows on the green axis.}
    \label{fig:circuit}
  \end{minipage}
\end{figure}

As illustrated in Figure \ref{fig:trace}, the black vectors represent the average intensity from each microphone, and the red vector shows their summation, indicating the estimated direction of the sound source. This method allows real-time directional estimation using minimal hardware and straightforward computations.

\subsection{Implementation}
The implemented sound-tracking system consists of three electret microphones arranged on a circular surface at equal angular intervals (0°, 120°, and 240°), each positioned 15 cm from the center. These microphones capture sound signals from different directions and are connected to LM358 operational amplifiers to boost the signal strength. The amplified outputs are fed into an Arduino Uno R3 microcontroller, which processes the signals and determines the direction of the incoming sound. A servo motor, also connected to the Arduino, rotates an indicator shaft toward the detected sound direction. Figure \ref{fig:circuit} illustrates the complete hardware setup, which is designed to be low-cost, efficient, and suitable for compact applications. The low sampling rate (50 Hz) necessitated placing the sound source close to the microphone array (35 cm) to ensure reliable direction estimation. While this limits the testing range, it demonstrates the feasibility of ultra-low-cost hardware.

\section{Results}
\subsection{Experimental Setup and Design}
To evaluate the performance of our sound-tracking system, we conducted 60 experiments using a 2-second audio clip played from a cellphone. The phone was positioned 35 cm from the center of the robot—close enough to ensure that all microphones received the signal, yet far enough to avoid saturation or directional bias. The phone was placed at different angles (20° and 120°) on a circular platform, and the system estimated the direction of the sound using the algorithm described earlier.

Each sound event triggered the collection of a fixed number of samples from the three microphones. The servo motor's directional tracking accuracy was then evaluated by converting the resulting vectors into rectangular coordinates using Equations \ref{sum} and \ref{sum2}. These were plotted as polar scatter plots (scatterpolar) to visually represent the direction and magnitude of the resulting vectors in (X,Y) format.

\subsection{Vector Visualization and Interpretation}
\begin{figure}[H]
  \centering
  \begin{minipage}{0.45\textwidth}
    \centering
    \includegraphics[scale=0.25]{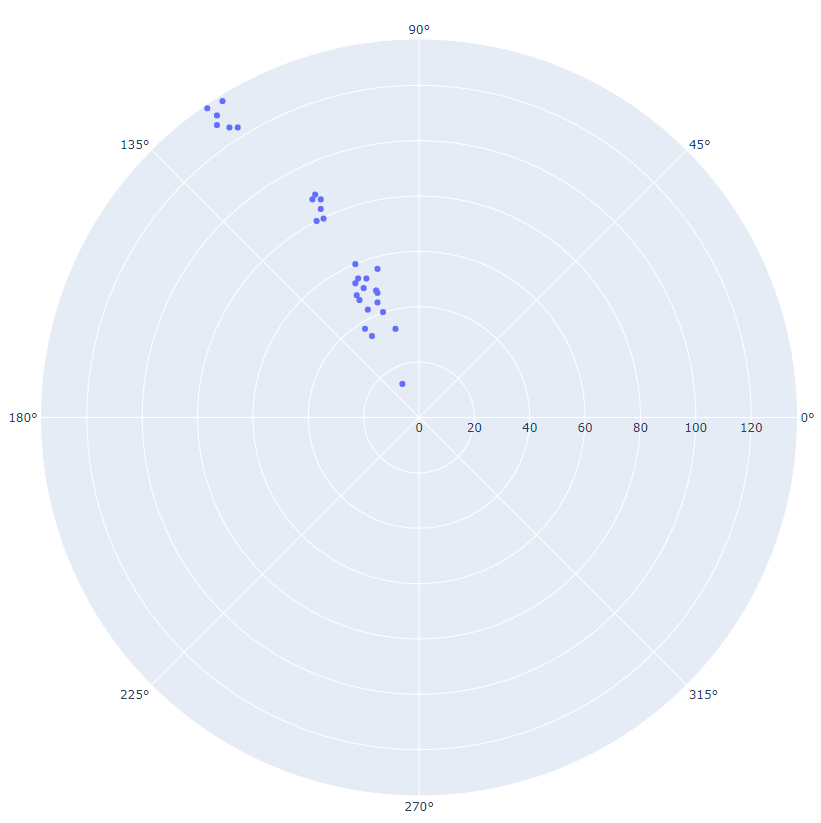}
    \caption{Detecting sound source angles and magnitude at 120 degrees}
    \label{fig:at120}
  \end{minipage}\hfill
  \begin{minipage}{0.45\textwidth}
    \centering
    \includegraphics[scale=0.25]{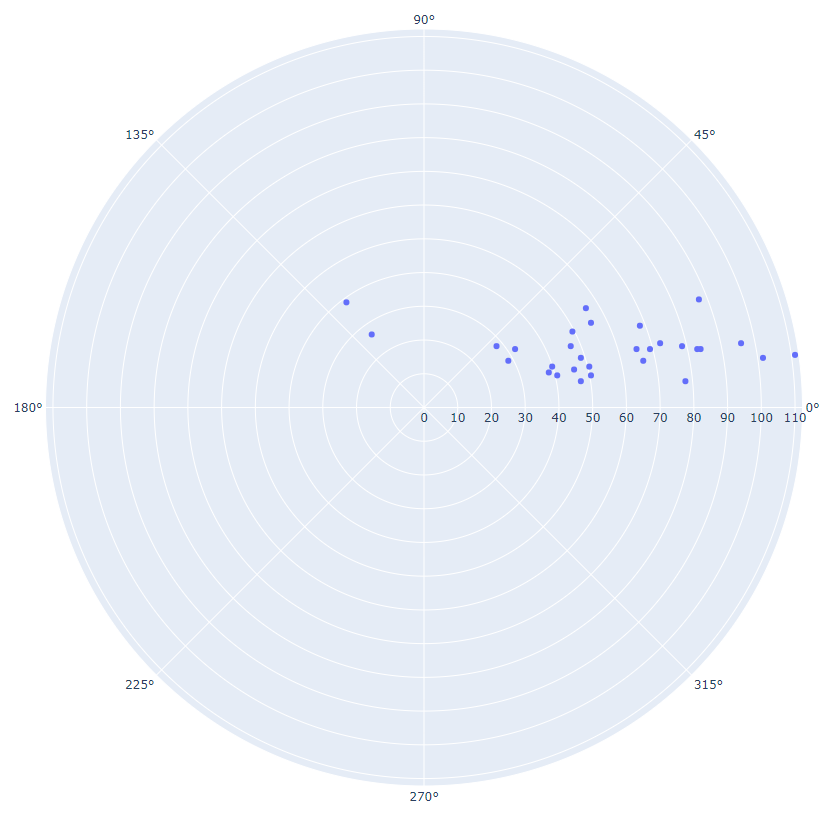}
    \caption{Detecting sound source angles and magnitude at 20 degrees}
    \label{fig:at20}
  \end{minipage}
\end{figure}

For each angle setting (120° and 20°), 30 tests were conducted. The vectors generated from each test were visualized in Figures \ref{fig:at120} and \ref{fig:at20}, respectively. Each dot in the scatterplot represents one test instance. The concentration of points along distinct directions indicates consistent directional estimation. Variations in the distance of points from the origin reflect differences in signal strength—stronger signals appearing farther from the center, and weaker ones appearing closer.

Variability in signal magnitude arises due to environmental noise and the inherent sensitivity differences among microphones. These factors contribute to minor deviations from ideal alignment in the plotted vectors but do not significantly affect the overall directional accuracy.

\subsection{Accuracy and Precision Analysis}
To ensure robustness, a 6\% trimming was applied to the sorted angle datasets, removing outliers and retaining 26 data points per target angle. At 120°, the system achieved an average accuracy of 5.26° with a precision of 3.26°, and at 20°, an accuracy of 7.11° with a precision of 4.01°.

These precision values were further interpreted in relation to a full 360° scale using the formula:
\[
\text{Precision} = \left(1 - \frac{\text{Precision in Degrees}}{360}\right) \times 100
\]
This corresponds to approximately 98.9\% and 98.8\% accuracy, respectively, reinforcing the high reliability of our approach.

\begin{table}[h]
    \centering
    \caption{Accuracy and Precision at Different Angles}
    \begin{tabular}{|l|c|c|} 
        \hline
        \textbf{Metric}            & \textbf{120 Degrees} & \textbf{20 Degrees} \\ \hline
        Number of Tests            & 30                   & 30                  \\ \hline
        Length of Trimmed Test      & 26                   & 26                  \\ \hline
        Accuracy (Degrees)         & 5.26                 & 7.11                \\ \hline
        Precision (Degrees)        & 3.26                 & 4.01                \\ \hline
        Precision Percentage       & 98.9\%              & 98.8\%             \\ \hline
    \end{tabular}
\end{table}

\section{Conclusion}

This paper introduced a simple and cost-effective method for sound-tracking using an Arduino, electret microphones, LM358 op-amps, and a servo motor. By applying a vector-sum approach, the system successfully estimated sound direction and controlled the servo to track the source. The results highlight the potential of this lightweight setup for practical applications in security, surveillance, robotics, and smart environments where directional audio is essential.

The proposed method demonstrates an affordable alternative to complex and expensive sound-tracking systems, making it accessible for real-time use in constrained environments. While effective, the experiments were limited by the Arduino Uno’s low sampling rate and memory capacity, requiring the sound source to remain close to the microphone array. This design choice reflects our focus on validating the algorithm’s feasibility on low-resource hardware rather than optimizing for large-scale deployment.

Future work will address these limitations by exploring higher sampling rates, more capable microcontrollers, and additional microphones or sensors to improve accuracy. Testing under realistic, noisy conditions will further evaluate robustness, extending the method’s applicability. Overall, this study contributes a practical foundation for affordable sound-tracking and opens opportunities for enhanced systems in robotics, automation, and acoustic sensing.

\bibliographystyle{splncs04}
\bibliography{references}


\end{document}